# Rethinking Structural Equation Modeling in Political Science: Challenges, Best Practices, and Future Directions

Bang Quan Zheng

University of Texas at Austin

2025

Abstract

Structural Equation Modeling (SEM) or Covariance Structure Analysis (CSA) is a versatile and powerful method in the social and behavioral sciences, providing a framework for modeling complex relationships, testing mediation, accounting for measurement error, and analyzing latent constructs. However, SEM remains underutilized in in political science; its application is often marred by misunderstandings, misinterpretations, and methodological pitfalls that can compromise the validity and interpretability of findings. This article examines key challenges in SEM applications within political science, including test statistics and fit indices, model specification, estimator selection, and causal inference. It offers practical recommendations for enhancing methodological rigor and introduces recent advancements in causal inference.

[Word count: 2,951]

# 1 Introduction

Structural Equation Modeling (SEM) or Covariance Structure Analysis (CSA) is a powerful tool in political science. While some view it as less contemporary compared to machine learning, network analysis, text-as-data, or other computational methods, its analytical power and versatility remain indispensable in political science, particularly in political psychology. SEM provides a rigorous framework for modeling complex relationships, testing mediation, analyzing latent constructs, and assessing measurement validity. More advanced applications—such as causal mediation analysis, measurement error correction, sensitivity analysis, multigroup confirmatory factor analysis (MG-CFA), and random intercept cross-lagged panel models (RI-CLPM)—can be integrated with modern causal inference techniques, further enhancing SEM's versatility.

Despite its broad utility and potential, SEM remains underutilized in political science, with its application often hindered by misunderstandings, misinterpretations, and methodological pitfalls that can compromise the validity and interpretability of findings. This paper provides a comprehensive overview of common challenges and best practices in SEM, identifies frequent errors in political science research, and offers practical guidance for improving its application while exploring its future direction in advancing causal inference.

# 2 Challenges in Applying SEM in Political Science

SEM is fundamentally distinct from traditional regression models, operating within its own statistical system. SEM relies on a strictly defined system of variance and covariance matrices that encompass both manifest and latent variables. A solid understanding of SEM requires a strong background in regression models, matrix algebra, and statistical inference. SEM evaluates the discrepancy between the sample covariance matrix and the model-implied covariance structure. For estimation to proceed, the variance and covariance matrix must be positive definite and invertible, a condition that is highly sensitive to factors such as the degrees of freedom—that is, the difference between data points and estimated parameters. Model convergence and fit are influenced by the ratio of the number of variables to the sample size. When this ratio is small, the eigenvalue structure of the sample covariance matrix will be ill-conditioned, rendering the covariance matrix noninvertible (Chi & Lange, 2014). Additionally, a small sample size relative to the degrees of freedom can inflate chi-square test statistics, increasing the risk of Type I error (Shi et al., 2019; Yuan & Bentler, 1999; Zheng & Bentler, 2021, 2023). Moreover, SEM is grounded in asymptotic statistical properties, where model identification, specification, estimation, and evaluation are deeply interconnected, requiring careful attention to each step to ensure valid and reliable results (Bentler, 1989; Bollen, 1989; Browne, 1982; Jöreskog & Sörbom, 1988). When these conditions are not met, models may suffer from non-convergence, improper solutions, inflated standard errors, or even negative variances. These common challenges can undermine result reliability, leading to biased estimates, and misleading conclusions.

However, most political science graduate programs do not include SEM in their methodological training. Many students attempt to self-teach SEM through online materials and by replicating sample code from *lavaan* (Rosseel, 2012) without fully grasping its complexities. Consequently, SEM applications in political science tend to be relatively basic, often limited to confirmatory factor analysis. What is more, political scientists have been slow to incorporate recent methodological advancements in SEM and other psychometric methods.

## 3 Best Practices for SEM in Political Science

### *3.1 Model Specification*

Model specification is crucial in SEM, as it impacts both parameter estimation—where key parameters may not be uniquely identified from the data—and model fit evaluation. However, political scientists often overlook this critical aspect. The most common issue is underspecification. Unlike psychology, which typically uses multiple items per construct, political science research often relies on too few indicators for latent variables. This results in limited degrees of freedom and an imbalanced ratio of data points to estimated parameters, leading to underspecified models. With too few indicators to adequately capture the latent constructs, model fit deteriorates, and estimation becomes unreliable.

The second major issue is model misspecification. Researchers construct a model based on prior knowledge and fit it to sample data by estimating its parameters. When modifications are needed, they adjust the model by adding or removing parameters and then refit it using the same dataset. If the initial model fit is poor, a common approach is to relax parameter constraints to improve alignment with the data (Bentler & Chou, 1992; MacCallum, Roznowski, & Nectowitz, 1992; Zheng and Bentler, 2025). This iterative process is known as model specification search.

While perfect model specification is unrealistic in real-world data analysis, a desirable model fit involves striking a balance between simplifying the model without compromising the overall fit and improving the model fit without making it more complicated (Bentler & Chou, 1992; MacCallum et al., 1992). A key symptom of misspecification is an excessively large chi-square test statistic, suggesting that the model fails to adequately capture the underlying data structure. This often occurs when researchers include too many free parameters without imposing sufficient constraints, such as fixing factor loadings or means. Without addressing identification issues, results become difficult to interpret. Additionally, omitted variables pose a significant challenge in SEM. If key latent variables are not adequately measured by manifest indicators, the model becomes misspecified.

The primary cause of model misspecification is omitted parameters or missing paths. To address this issue, researchers should conduct specification searches and apply appropriate model modifications. The Lagrange Multiplier (LM) test, introduced by Lee and Bentler (1980), offers a forward stepwise approach to detect omitted parameters, while the Wald test provides a backward stepwise approach. Both tests are available in *lavaan, Mplus*, or other SEM software. To illustrate, researchers should assess model performance under misspecification and compare it with a revised model based on the LM or Wald test to ensure robustness and validity.

## *3.2 SEM Estimation*

Selecting the correct estimator based on data properties is essential for ensuring accurate model estimation and valid statistical inferences in SEM. Misestimations often lead to biased coefficients and misleading conclusions. The default estimator in most SEM software, including *lavaan* and *Mplus*, is maximum likelihood, which assumes multivariate normality and requires a large sample size with normally distributed data for optimal performance. However, these assumptions are frequently violated in real-world data analysis, often resulting in inflated chi-square values, increasing the risk of Type I error. Therefore, it is crucial to check data normality, assess variable characteristics, and choose the appropriate estimator to address these issues effectively.

Three key estimators should be considered based on data characteristics, distribution, and sample size:

1. Non-Normal Data: When the data deviates from normality, the Satorra-Bentler test provides a robust alternative. The chi-square test statistic adjusted using the Satorra-Bentler scaled chi-square remains remarkably stable across a wide range of sample sizes (Zheng & Bentler, 2024).
2. Small Sample Size: If the sample size is small (e.g., fewer than 200 cases), using the reweighted least squares (RLS) estimator can provide more stable chi-square test estimates, improving model reliability (Hayakawa, 2019; Zheng & Bentler, 2021).

3. Categorical Data: When working with categorical indicators, ML estimation is inappropriate. Instead, estimators such as weighted least squares mean and variance adjusted (WLSMV) should be used to handle non-continuous variables effectively.

## *3.3 Chi-Square Test Statistics and Fit Indices*

Model fit in SEM is often complex and challenging to interpret, as it depends on assessing the discrepancy between the model-implied covariance structure and the sample covariance structure. Ideally, the chi-square test statistic should be close to the model's degrees of freedom and yield a p-value greater than 0.05, allowing researchers to retain the null hypothesis that the model adequately fits the data. However, in a misspecified model, the model-implied covariance deviates from the sample covariance, leading to inflated chi-square test statistics and a p-value near zero.

The core of SEM lies in evaluating model fit, which primarily depends on chi-square test statistics and various fit indices, such as the normed fit index (NFI), comparative fit index (CFI), Tucker-Lewis index (TLI), and root mean square error of approximation (RMSEA). However, no clear guidelines exist for integrating chi-square test statistics with fit indices when interpreting model fit. Scholars continue to debate the validity and applicability of fit indices, given the lack of consensus on which ones to prioritize.

Some researchers argue that fit indices have limited practical value (Barrett, 2007) and advocate for relying solely on the chi-square statistic, warning that fit indices can lead researchers to falsely conclude that a misspecified model is adequate (Stone, 2021). Others caution against rigidly applying cutoff values for fit indices, as these thresholds can be misleading and prone to misuse. A related concern is "cherry-picking," where researchers selectively report fit indices that support their hypotheses, thereby justifying a poorly fitting model (Jackson et al., 2009; Kline, 2015; Stone, 2021).

Fit indices such as RMSEA, NFI, CFI, and TLI are widely used in political science SEM applications but are often misinterpreted. Many researchers rigidly adhere to cutoff criteria without considering model complexity, sample size, or theoretical justification. Additionally, omitting a comprehensive set of fit indices can create a false sense of confidence in poorly specified models. Using the Monte Carlo simulations, Zheng and Bentler (2025) demonstrate that fit indices are unreliable for assessing model fit, as they perform poorly in cases of model misspecification, small sample sizes, and non-normal data.

To improve SEM applications, scholars should adopt a more rigorous approach to model evaluation and estimation. First, researchers must go beyond fit indices and integrate chi-square test statistics into their assessments. Scholars advocate that the chi-square test should remain the primary model fit metric, as all other fit indices are derived from it (Zheng & Bentler, 2024). It is advisable to always report chi-square test statistics and fit indices. Even when chi-square values are large with a p-value close to zero, reporting them remains essential for assessing model adequacy. A key rule of thumb is to compare chi-square test statistics with degrees of freedom: 1) If the chi-square statistic is only slightly or moderately larger than the degrees of freedom, the model might suffer from minor underspecification or misspecification. In this case, fit indices such as NFI, CFI, and RMSEA can be important supporting information for model fit. 2) If the chi-square statistic substantially exceeds the degrees of freedom, the model is likely severely misspecified, leading to biased fit indices. Overall, properly evaluating model fit requires a holistic approach, incorporating both chi-square test statistics and fit indices, while being mindful of sample size, model complexity, and theoretical justification to avoid misleading conclusions.

## 4 Expanding SEM's Role in Causal Inference

Causal inference is increasingly important in political science, and SEM provides a powerful framework for analyzing causal relationships. While SEM models complex pathways, causality is only established when supported by strong theory, experimental designs, or instrumental variables. In experimental research, multi-group confirmatory factor analysis (MG-CFA) enables latent construct analysis within treatment and control groups. Establishing configural, metric, and scalar invariance ensures consistent interpretation of measurement items, allowing valid causal inferences. For example, in a political psychology experiment on media exposure and political trust, researchers can measure trust in government using multiple survey items. Achieving invariance ensures that observed differences result from the treatment rather than variations in measurement interpretation.

Second, mediation analysis, with direct and indirect effects as fundamental components of the SEM framework, has long been a key analytical approach (Cheong & MacKinnon, 2012; MacKinnon, 2008). In recent years, causal mediation analysis (CMA) has gained increasing adoption for investigating causal mechanisms—an approach fundamentally rooted in SEM. The rise of CMA in political science is largely attributed to the seminal work of Imai et al. (2010) and Imai et al. (2011). They streamlined traditional mediation effects by relaxing parametric assumptions, making the approach more adaptable to political science applications within both parametric and non-parametric frameworks. Additionally, sensitivity analysis in causal mediation analysis is also grounded in SEM, as it helps assess the robustness of findings by addressing potential violations of the sequential ignorability assumption, which underlies the relationship between the treatment and the dependent variable. The growing use of CMA and sensitivity analysis further highlights SEM's critical role in advancing causal inference in political science.

Third, for longitudinal causal analysis, scholars have traditionally relied on Cross-Lagged Panel Models (CLPM) to support causal claims, as it is widely recognized as a standard method in political science. However, its application is often susceptible to unaccounted time-invariant confounders and measurement error, which can lead to biased estimates of autoregressive and cross-lagged relationships, as well as incorrect statistical inferences and substantive conclusions. More recently, the development of Random Intercept Cross-Lagged Panel Model (RI-CLPM) has allowed researchers to more effectively address measurement error and unobserved confounding by accounting for stable trait variance while isolating within-person variance, thus capturing intra-individual change (Hamaker et al., 2015; Usami et al., 2019). These advanced techniques extend well beyond the methodological toolbox of most political scientists, underscoring the need for greater familiarity with SEM principles.

Many political scientists struggle with complex SEM models like MG-CFA, CMA, CLPM, and RI-CLPM due to their learning and interpretation challenges. A strong foundation in SEM is essential for proper model specification, estimator selection, and model fit assessment. While SEM does not inherently establish causality, a deep understanding ensures its effective adaptation and application for rigorous analysis.

## 5 Conclusion

Political methodology continues to evolve, yet SEM remains a powerful but underutilized tool in political science, often hindered by misunderstandings and methodological challenges. To maximize its potential, scholars must move beyond basic confirmatory factor analysis and develop a deeper understanding of model specification, estimation techniques, and chi-square test statistics. While new methods like text-as-data and network analysis are gaining traction, SEM's versatility—especially in political psychology—makes it indispensable for studying complex relationships. Expanding its application to causal inference through techniques like causal mediation analysis and random intercept cross-lagged panel models can yield more robust and theoretically meaningful insights. By refining SEM practices and fully leveraging its strengths, political scientists can enhance research on political behavior and institutions.

# References

Barrett, P. (2007). Structural Equation Modeling: Adjudging Model Fit. *Personality and Individual Differences*, *42*(815-824).

Bentler, Peter. (1989). *EQS Structural Equations Program Manual*. BMDP Statistical Software.

Bentler, Peter M., & Chou Chih-Ping. (1992). Some New Covariance Structure Model Improvement Statistics. *Sociological Methods & Research*, *21*(2), 259-282. https://doi.org/https://doi.org/10.1177/0049124192021002006

Bollen, Kenneth A. (1989). *Structural Equations with Latent Variables*. John Wiley & Sons.

Browne, Michael. (1982). Covariance structures. In D. M. Hawkins (Ed.), *Topics in applied multivariate analysis* (pp. 72-141). Cambridge University Press.

Cheong, J., & Mackinnon D. P. (2012). Mediation/Indirect Effects in Structural Equation Modeling. In R. H. Hoyle (Ed.), *Handbook of structural equation modeling* (pp. 417-435). The Guilford Press.

Chi, Eric C., & Lange Kenneth. (2014). Stable estimation of a covariance matrix guided by nuclear norm penalties. *Computional Statistics and Data Analysis*(80), 117-128.

Hamaker, E. L., Kuiper R. M., & Grasman R. P. P. P. (2015). A critique of the cross-lagged panel model. *Psychological Methods*, *20*, 102-116.

Hayakawa, Kazuhiko. (2019). Corrected Goodness-of-fit test in covariance structure analysis. *Psychological Methods*, *24*(3), 371-389.

Imai, K., Keele L., & Tingley D. (2010). A General Approach to Causal Mediation Analysis. *Psychol Methods*, *15*(4), 309-326.

Imai, K., Keele L., Tingley D., & Yamada T. (2011). Unpacking the Black Box of Causality: Learning about Causal Mechanisms from Experimental and Observational Studies. *American Political Science Review*, *105*(4), 765789.

Jackson, D. L., Gillaspy A. J., & Purc-Stephenson R. (2009). Reporting Practices in Confirmatory Factor Analysis: An Overview and Some Recommendations. *Psychological Methods*, *14*(1), 6-23. https://doi.org/10.1037/a0014694

Jöreskog, Karl Gustav, & Sörbom Dag. (1988). *LISREL 7, A Guide to the Program and Applications*. SPSS.

Kline, Rex B. . (2015). *Principles and Practice of Structural Equation Modeling*. Guilford publications.

Lee, Sik-Yum, & Bentler Peter. (1980). Some Asymptotic Properties of Constrained Generalized Least Squares Estimation in Covariance Structure Models. *South African Statistical Journal*, *14*, 121-136.

Maccallum, Robert C., Roznowski Mary, & Nectowitz Lawrence B. (1992). Model Modifications Covariance Structure Analysis: The Problem of Capitalization on Chance. *Psychological Bulletin*, *111*(3), 490-504.

Mackinnon, D. P. (2008). *Introduction to Statistical Mediation Analysis*. Routledge.

Rosseel, Yves. (2012). lavaan: An R package for structural equation modeling. *Journal of Statistical Software*, *48*(2), 1-36.

Shi, D., Lee T., & Maydeu-Olivares A. (2019). Understanding the model size effect on SEM fit indices. *Educational and Psychological Measurement*, *79*(2), 310-334.

Stone, Bryant. (2021). The Ethnical Use of Fit Indices in Structural Equation Modeling: Recommendations for Psychologists. *Frontiers in Psychology*, *12*(782226).

Usami, Satoshi, Murayama K, & Hamaker E. L. . (2019). A unified framework of longitudinal models to examine reciprocal relations. *Psychol Methods*, *24*(5), 637-657.

Yuan, K.H., & Bentler Peter. (1999). On Asymptotic Distribution s of Normal Theory MLE in Covariance Structure Analysis Under Some Nonnormal Distributions. *Statistics and probability Letters*(42), 107-113.

Zheng, Bang Quan, & Bentler Peter M. (2021). Testing Mean and Covariance Structures with Reweighted Least Squares. *Structural Equation Modeling: A Multidisplinary Journal*. https://doi.org/https://doi.org/10.1080/10705511.2021.1977649

Zheng, Bang Quan, & Bentler Peter M. (2023). RGLS and RLS in Covariance Structure Analysis. *Structural Equation Modeling: A Multidisplinary Journal*, *30*(2), 234-244. https://doi.org/https://doi.org/10.1080/10705511.2022.2117182

Zheng, Bang Quan, & Bentler Peter M. (2024). Enhancing Model Fit Evaluation in SEM: Practical Tips for Optimizing Chi-Square Tests. *Structural Equation Modeling: A Multidisplinary Journal*, *32*(1), 136-141. https://doi.org/https://doi.org/10.1080/10705511.2024.2354802